\documentstyle[aps]{revtex}

\input epsf

\begin{document}
\title{Pinning enhancement upon the magnetic flux trapping in the clusters of a
normal phase with fractal boundaries}
\author{Yuriy I. Kuzmin}
\address{Ioffe Physical Technical Institute of the Russian Academy of Sciences,\\
Polytechnicheskaya 26 St., Saint Petersburg 194021 Russia,\\
and State Electrotechnical University of Saint Petersburg,\\
Professor Popov 5 St., Saint Petersburg 197376 Russia\\
e-mail: yurk@mail.ioffe.ru; iourk@yandex.ru\\
tel.: +7 812 2479902; fax: +7 812 2471017}
\date{\today}
\maketitle

\begin{abstract}
The magnetic flux trapping in type-II superconductor containing fractal
clusters of a normal phase, which act as pinning centers, is considered. The
critical current distribution for an arbitrary fractal dimension of the
boundaries of the normal phase clusters is obtained. It is revealed that the
fractality of the cluster boundaries intensifies the pinning and thereby
raises the critical current of a superconductor. The pinning gain
coefficient as a function of critical current for different fractal
dimensions is calculated.
\end{abstract}

\bigskip

Isolated clusters of a normal phase in superconductors can act as effective
pinning centers \cite{hig}-\cite{dor}. The presence of such clusters may
significantly affect the vortex dynamics and transport properties,
especially when the clusters have fractal boundaries \cite{lai}-\cite{pla}.

Let us consider a superconducting film containing columnar inclusions of a
normal phase, which are oriented transversely to the film surface. When the
specimen is cooled below the critical temperature in the magnetic field
directed along the axis of the normal phase inclusions, the magnetic flux
will be frozen in the isolated clusters of a normal phase so the space
distribution of the trapped magnetic field will be two--dimensional. Let us
suppose that the film surface fraction covered by the normal phase is below
the percolation threshold for the transfer of magnetic flux. Then there is a
superconducting percolation cluster in the plane of the film where a
transport current can flow. When the current increases the trapped flux
remains unchanged until the vortices start to break away from the clusters
of pinning force weaker than the Lorentz force created by the transport
current. As this happens, the vortices have to cross the surrounding
superconducting space and they will first do that through the weak links,
which connect the adjacent normal phase clusters between themselves. In
high-temperature superconductors (HTS), which are characterized by an
extremely short coherence length, such weak links form readily in sites of
various structural defects \cite{kup}-\cite{has}. Moreover, a magnetic field
further reduces the coherence length \cite{son}, thus resulting in more easy
weak links formation. In conventional low-temperature superconductors, which
are characterized by a large coherence length, weak links can be formed due
to the proximity effects in sites of minimum distance between the next
normal phase clusters.

According to the local configuration of weak links each normal phase cluster
has its own value of the critical current, which contributes to the overall
statistical distribution. By the critical current of the cluster we mean the
current of depinning, that is to say, such a current at which the magnetic
flux ceases to be held inside the cluster of a normal phase.\ When a
transport current is gradually increases, the vortices will break away first
from clusters of small pinning force and, therefore, of small critical
current. Thus the change in the trapped magnetic flux $\Delta \Phi $ caused
by the action of the transport current $I$ is proportional to the total
number of the normal phase clusters of critical currents less than a preset
value $I$. Therefore, the relative change in the magnetic flux can be
expressed with the cumulative probability function $F=F\left( I\right) $ for
the distribution of the critical currents
\begin{equation}
\frac{\Delta \Phi }{\Phi }=F\left( I\right) \text{, \ \ \ \ \ \ \ \ \ \ \ \
\ \ \ \ \ \ \ \ \ \ \ \ \ \ \ }F\left( I\right) =\Pr \left\{ \forall
I_{j}<I\right\} \text{ \ ,}  \label{probi}
\end{equation}
where $\Pr \left\{ \forall I_{j}<I\right\} $ is the probability that any $j$%
th cluster has a critical current $I_{j}$ less than a given upper bound $I$.

At the same time, the larger the normal phase cluster size, the more weak
links are along its border with the surrounding superconducting space, and
therefore, the smaller is the critical current of this cluster. If the
concentration of entry points into weak links\ per unit perimeter length is
constant for all clusters regardless of their size, the critical current $I$%
\ is inversely proportional to the cluster perimeter length $P$: $I\propto
1/P$. Taking into account that the magnetic flux trapped in a cluster is
proportional to its area $A$, the relative change in the total trapped flux
can be expressed with the cumulative probability function $W=W\left(
A\right) $ describing the distribution of the areas of the normal phase
clusters:
\begin{equation}
\frac{\Delta \Phi }{\Phi }=1-W\left( A\right) \text{, \ \ \ \ \ \ \ \ \ \ \
\ \ \ \ \ \ \ \ \ \ \ \ \ \ \ \ }W\left( A\right) =\Pr \left\{ \forall
A_{j}<A\right\} \text{ \ ,}  \label{proba}
\end{equation}
where $\Pr \left\{ \forall A_{j}<A\right\} $ is the probability that the
area $A_{j}$ of an arbitrary $j$th cluster does not exceed a given upper
bound $A$. The distribution function $W=W\left( A\right) $ of the cluster
areas can be found by a geometric probability analysis of electron
photomicrographs of superconducting films \cite{tpl}. Thus in the
practically important case of YBCO films with columnar defects \cite{pss}
the exponential distribution of the cluster areas can be realized:
\begin{equation}
W\left( A\right) =1-\exp \left( \frac{A}{\overline{A}}\right)  \label{expo}
\end{equation}
where $\overline{A}$ is the mean area of the cluster.

Thus, in order to clear up how the transport current acts on the trapped
magnetic flux, it is necessary to find out the relationship between the
distribution of the critical currents [Eq.~(\ref{probi})] and areas [Eq.~(%
\ref{proba})] of the normal phase clusters. As was first found in Ref.~\cite
{pla}, clusters of a normal phase can have fractal boundaries, and this
feature significantly affects the dynamics of the trapped magnetic flux. For
fractal clusters the perimeter-area relation has the form
\begin{equation}
P\propto A^{D/2}  \label{scaling}
\end{equation}
which leads us to the following expression for the critical current of the
cluster: $I=\alpha A^{-D/2}$, where $\alpha $ is the form factor and $D$ is
the fractal dimension of the cluster boundary.

The relation of Eq.~(\ref{scaling}) is consistent with the generalized
Euclid theorem \cite{man82}, which states that the ratios of the
corresponding measures are equal when reduced to the same dimension. So it
means that $P^{1/D}\propto A^{1/2}$, which is valid both for Euclidean ($D=1$%
) and fractal ($D>1$) geometric objects. Inasmuch as the cluster boundary is
a fractal, it is the statistical distribution of the cluster areas, rather
than their perimeters, that is fundamental for finding the critical current
distribution. The topological dimension of the cluster perimeter (equal to
unity) does not coincide with its Hausdorff-Bezikovich dimension, which
strictly exceeds unity. Therefore the perimeter length of a fractal cluster
is not well defined, because its value depends on the resolution capacity of
the cluster geometric size measurement. At the same time, the topological
dimension of the cluster area is the same as the Hausdorff-Bezikovich one
(both are equal to 2). Thus, the area restricted by the fractal curve is a
well-defined finite quantity.

Speaking of the geometric characteristics of the normal phase clusters, we
are considering the cross-sections of the extended objects, which indeed the
normal phase clusters are, by the plane carrying a transport current.
Therefore, though a normal phase cluster represents a self-affine fractal
\cite{man86}, we will analyze its geometric probability properties in the
planar section only, where the boundary of the cluster is statistically
self-similar.

In accordance with starting formulas of Eq.~(\ref{probi}) and Eq.~(\ref
{proba}) the exponential distribution of the cluster areas of Eq.~(\ref{expo}%
) gives rise to an exponential-hyperbolic distribution of critical currents:
\begin{equation}
F\left( i\right) =\exp \left( -\left( \frac{2+D}{2}\right)
^{2/D+1}i^{-2/D}\right)  \label{exhyp}
\end{equation}
where $i\equiv I/I_{c}$ is the dimensionless transport current and $%
I_{c}=\left( 2/\left( 2+D\right) \right) ^{\left( 2+D\right) /2}\alpha
\left( \overline{A}\right) ^{-D/2}$ is the critical current of the
transition into a resistive state.

\begin{figure}[tbp]
\epsfbox{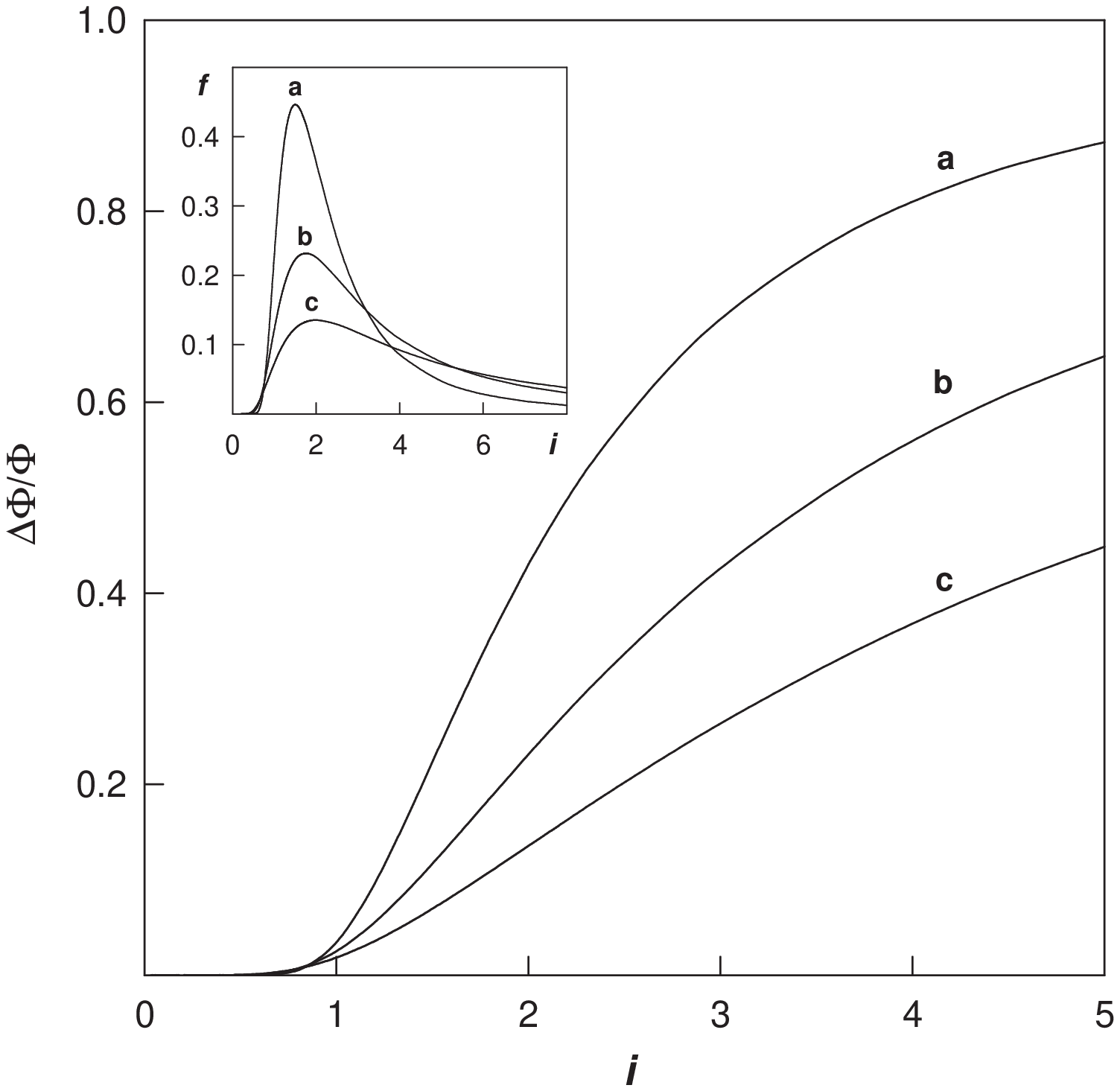}
\caption{Effect of a transport current on
trapped magnetic flux at different
values of the fractal dimension. Curve ({\sf a}) corresponds to the case of $%
D=1$; curve ({\sf b}) of $D=1.5$; curve (${\sf c}$) of $D=2$. The inset
shows the corresponding critical current distributions.}
\label{figure1}
\end{figure}

The distribution function of Eq.~(\ref{exhyp}) allows us to fully describe
the effect of the transport current on the trapped magnetic flux (see Fig.~%
\ref{figure1}). Knowing the cumulative probability function, the probability
density $f(i)\equiv dF/di$ for the critical current distribution can be
readily derived:
\[
f\left( i\right) =\frac{2}{D}\left( \frac{2+D}{2}\right)
^{2/D+1}i^{-2/D-1}\exp \left( -\left( \frac{2+D}{2}\right)
^{2/D+1}i^{-2/D}\right)
\]

The relative change in the trapped magnetic flux $\Delta \Phi /\Phi $, which
can be found from Eq.~(\ref{exhyp}), is proportional to the density of
vortices $n$ broken away from the pinning centers by the current $i$:
\[
n\left( i\right) =\frac{B}{\Phi _{0}}\int_{0}^{i}f\left( i^{\prime }\right)
di^{\prime }=\frac{B}{\Phi _{0}}\frac{\Delta \Phi }{\Phi }\text{,}
\]
where $B$ is the magnetic field and $\Phi _{0}\equiv hc/\left(
2e\right) $ is the magnetic flux quantum ($h$ is Planck's
constant, $c$ is the velocity of light, and $e$ is the electron
charge). As may be seen from Fig.~\ref {figure1}, breaking of the
vortices away from the pinning centers becomes significant only
for $i>1$, i.e. after a resistive transition. Up to this point the
trapped flux remains virtually unchanged because the Lorentz force
created by such a small current is weaker than the pinning force.

Figure \ref{figure1} shows how the fractality of the cluster boundary
affects the magnetic flux trapping. Graph ({\sf a}) corresponds to extreme
case of Euclidean clusters ($D=1$), while graph ({\sf c}) relates to the
clusters of boundaries with maximum fractality ($D=2$; as an example, the
Peano curves have such a fractal dimension.) Whatever the geometry and
morphology of the clusters may be, their graphs \{$\Delta \Phi /\Phi $ vs $%
i\}$ will always fall within the region between these two limiting curves.
As an example, graph ({\sf b}) for the case of fractal dimension $D=1.5$ is
shown.

Figure \ref{figure1} demonstrates an important consequence of
Eq.~(\ref {exhyp}), according to which the fractality of normal
phase clusters hinders the breaking away of the vortices and
thereby strengthens pinning. As may be clearly seen from the inset
of Fig.~\ref{figure1}, the bell--shaped curve of the critical
current distribution broadens out, moving towards greater
magnitudes of current as the fractal dimension increases. The
pinning amplification due to the fractality can be characterized
by the pinning gain factor
\[
k_{\Phi }\equiv 20\log \frac{\Delta \Phi \left( D=1\right)
}{\Delta \Phi \left( \text{current value of }D\right) }\text{ , \
\ \ \ \ dB}
\]
which is equal to the relative decrease in the fraction of magnetic flux
broken away from fractal clusters of the dimension $D$ compared to the case
of Euclidean ones ($D=1$). This quantity can be calculated from to following
formula:
\[
k_{\Phi }=\frac{20}{\ln 10}\left( \left( \frac{2+D}{2}\right)
^{2/D+1}i^{-2/D}-\frac{3.375}{i^{2}}\right)
\]

\begin{figure}[tbp]
\epsfbox{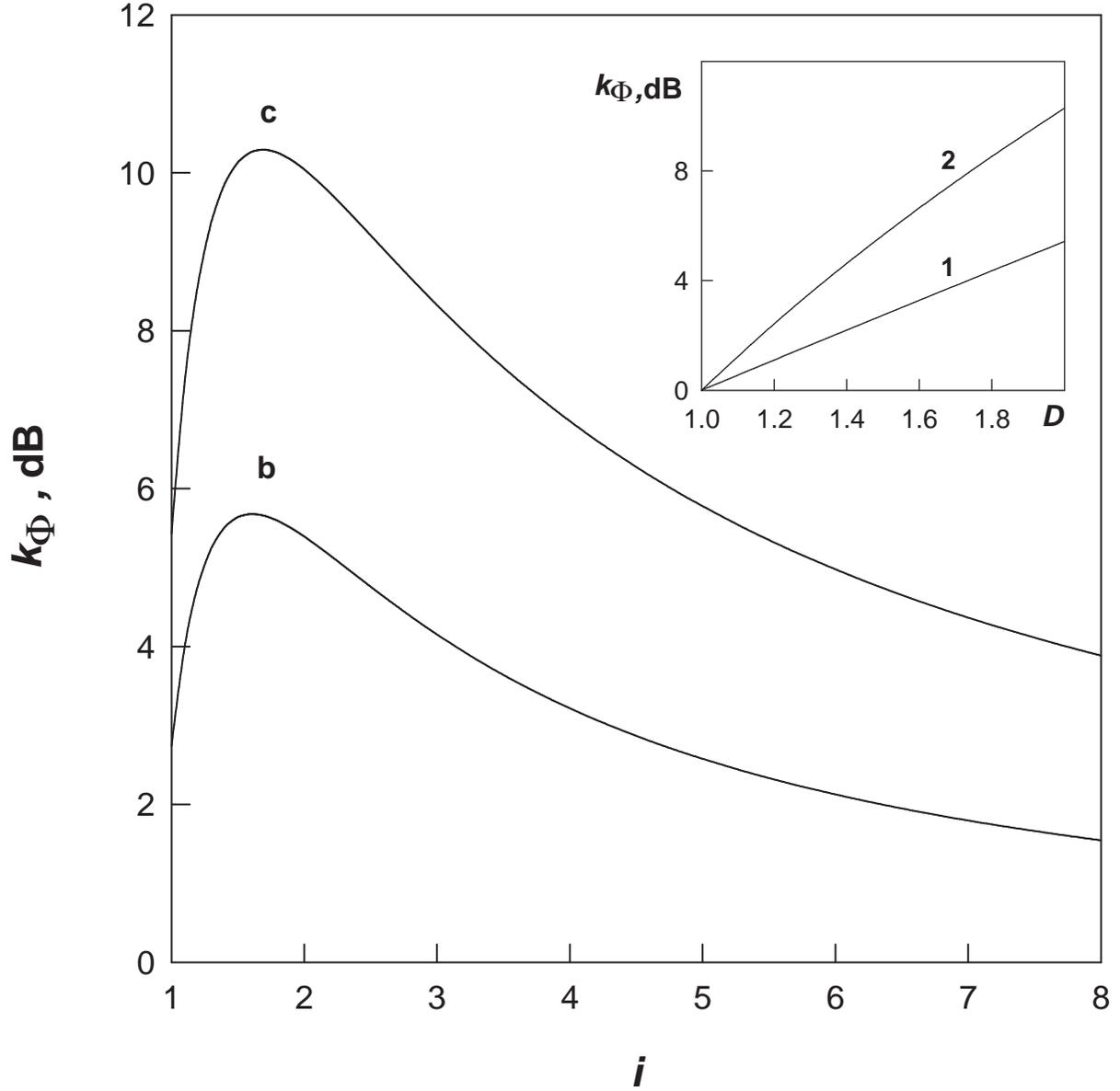}
\caption{Pinning gain as a function of
transport current at different values of the fractal dimension:
$D=1.5$ ({\sf b}) and $D=2$ (${\sf c}$). The inset shows the
dependence of the pinning gain on the fractal dimension at fixed
values of a transport current: $i=1$ (curve (1)) and $i=1.6875$
(here the pinning gain has a maximum, curve (2)).} \label{figure2}
\end{figure}

The dependences of the pinning gain on the transport current as well as on
the fractal dimension are given in Fig.~\ref{figure2}. The highest
amplification is reached when the clusters have the maximum fractality ($D=2$%
):
\[
\mathrel{\mathop{\max }\limits_{D}}%
k_{\Phi }=\frac{20}{\ln 10}\left( \frac{4i-3.375}{i^{2}}\right)
\]
Let us note that the pinning gain characterizes the transport
properties of a superconductor in the range of the transport
currents corresponding to a resistive state ($i<1$). At smaller
currents the total trapped flux remains unchanged (see
Fig.~\ref{figure1}) for a lack of pinning centers of such small
critical currents, so the breaking away of the vortices has not
started yet. When the vortices leave the normal phase clusters and
start to move, their motion induces an electric field, which, in
turn, creates a voltage drop across the sample. Thus, at a
transport current greater than the current of the resistive
transition some finite resistance appears, so that the passage of
electric current is accompanied by energy dissipation. As for any
hard superconductor (type--II, with pinning centers) this
dissipation does not mean the destruction of phase coherence yet.
Some dissipation always accompanies any motion of a magnetic flux
that can happen in a hard superconductor even at low transport
current. Therefore the critical current in such materials cannot
be specified as the greatest non-dissipative current. The
superconducting state collapses only when the growth of
dissipation becomes avalanche-like as a result of thermomagnetic
instability.

Thus, the fractality of normal phase cluster\ boundaries strengthens the\
magnetic flux pinning. This phenomenon provides principally new
possibilities for increasing the current-carrying capability of composite
superconductors by optimizing their geometric morphological properties with
no changes of the chemical constitution.\bigskip

\begin{center}
${\bf Acknowledgements\smallskip }$
\end{center}

This work is supported by the Russian Foundation for Basic Researches (Grant
No 02-02-17667).

\newpage

\end{document}